\documentclass[conference]{IEEEtran}
\usepackage{fancyhdr}
\usepackage{balance}
\usepackage{amsmath}
\usepackage{color}
\usepackage{soul}
\usepackage{mathrsfs}
\usepackage{amssymb}
\usepackage{graphicx}
\usepackage{amsfonts}
\usepackage{booktabs}
\usepackage{algorithm} 
\usepackage{algorithmic} 

\DeclareMathOperator*{\argmax}{arg\,max}

\begin{document}
%
\title{Structured Matching Pursuit for Reconstruction of Dynamic Sparse Channels}
\author{\IEEEauthorblockN{Xudong Zhu${}^{1}$, Linglong Dai${}^{1}$, Guan Gui${}^{2}$, Wei Dai${}^{3}$, Zhaocheng Wang${}^{1}$ and Fumiyuki Adachi${}^{4}$}
\IEEEauthorblockA{
${}^{1}$Tsinghua National Laboratory for Information Science and Technology (TNlist), Tsinghua University, China\\
${}^{2}$Department of Electronics and Information Systems, Akita Prefectural University, Japan\\
${}^{3}$Department of Electrical and Electronic Engineering, Imperial College, UK\\
${}^{4}$Department of Communications Engineering, Tohoku University, Japan\\
Email: daill@tsinghua.edu.cn, wei.dai1@imperial.ac.uk, adachi@ecei.tohoku.ac.jp}}

\IEEEoverridecommandlockouts


\maketitle

\begin{abstract}
In this paper, by exploiting the special features of temporal correlations of dynamic sparse channels that path delays change slowly over time but path gains evolve faster, we propose the structured matching pursuit (SMP) algorithm to realize the reconstruction of dynamic sparse channels.
Specifically, the SMP algorithm divides the path delays of dynamic sparse channels into two different parts to be considered separately, i.e., the common channel taps and the dynamic channel taps.
Based on this separation, the proposed SMP algorithm simultaneously detects the common channel taps of dynamic sparse channels in all time slots at first, and then tracks the dynamic channel taps in each single time slot individually.
Theoretical analysis of the proposed SMP algorithm provides a guarantee that the common channel taps can be successfully detected with a high probability, and the reconstruction distortion of dynamic sparse channels is linearly upper bounded by the noise power.
Simulation results demonstrate that the proposed SMP algorithm has excellent reconstruction performance with competitive computational complexity compared with conventional reconstruction algorithms.
\end{abstract}

\IEEEpeerreviewmaketitle

\section{Introduction}
In broadband wireless communication systems, channel state information is required at the receiver for coherent signal detection since wireless channel distorts the received signals, especially when the wireless channel is dynamically changing.
Therefore, accurate channel estimation becomes a fundamental problem for broadband wireless communication systems over dynamic wireless channels \cite{CEimportant}.

Various linear channel estimation methods have been proposed in the literature \cite{CEhistory}, but their performance can hardly meet the simultaneously booming demand of high-rate and high-mobility wireless communications.
Recently, a lot of physical channel measurements have verified that wireless channels exhibit sparsity, i.e., the dimension of a wireless channel may be large, but the number of channel taps with significant power is usually small, especially in broadband wireless communication systems \cite{channelsparsity}.
By exploiting this channel sparsity, compressive sensing (CS) algorithms have been used to improve the channel reconstruction performance, such as orthogonal matching pursuit (OMP) \cite{OMP}, compressive sampling matching pursuit (CoSaMP) \cite{CoSaMP}, subspace pursuit (SP) \cite{SP}, etc.
Compared with conventional linear methods, CS-based channel reconstruction methods are able to achieve reliable channel estimation with reduced training resources \cite{GuanGui}.

Further studies have uncovered additional channel characteristics, e.g., the temporal correlations of practical wireless channels: path delays change slowly over time, while path gains evolve faster \cite{dynamicchannel}.
By taking these temporal channel correlations into account, we have proposed several simultaneous CS-based methods to realize simultaneous reconstruction of sparse channels, such as adaptive simultaneous OMP (A-SOMP) \cite{ASOMP} and structured SP (SSP) \cite{SSP} algorithms.
Different from the CS-based methods ignoring the temporal correlations of dynamic sparse channels but reconstructing the channel in each time slot independently, the simultaneous CS-based methods assume that dynamic channels in several consecutive time slots share the same path delay set to improve the channel estimation performance.
However, the path delays of dynamic sparse channels may change over time, or even there maybe some mutations of the path delays, so the assumption of the simultaneous CS-based methods is not always true in practical broadband wireless communication systems \cite{dynamicchannel}.

In this paper, we propose a structured CS algorithm called structured matching pursuit (SMP) for the reconstruction of dynamic sparse channels in broadband wireless communication systems.
By exploiting the temporal correlations of a dynamic sparse channel \cite{dynamicchannel}, we divide the path delays of the dynamic sparse channel into two different parts to be considered separately, i.e., the common channel taps and the dynamic channel taps.
Based on this separation, the proposed SMP algorithm simultaneously detects the common channel taps of the dynamic sparse channel in all time slots at first.
Then the path delay set of the common channel taps is used as the initialization set in the dynamic channel taps tracking process, which aims to track the dynamic channel taps and remove the fake taps in the initialization set in each time slot.
The theoretical analysis of the proposed SMP algorithm based on the tool of restricted isometry property (RIP) indicates that the common channel taps can be successfully detected with a high probability, and the reconstruction distortion of the dynamic sparse channel is linearly upper bounded by the noise power.
Numerical simulations show that the proposed SMP algorithm with competitive computational complexity has better channel reconstruction performance than conventional reconstruction algorithms.

The rest of this paper is organized as follows.
Section II presents the system model.
Section III addresses the proposed SMP algorithm.
Section IV presents the performance analysis, and simulation results are provided in Section V.
Finally, conclusions are drawn in Section VI.

\emph{Notation}:
We use upper-case and lower-case boldface letters to denote matrices and vectors, respectively;
$(\cdot)^T$, $(\cdot)^H$, $(\cdot)^{-1}$, $(\cdot)^{\dagger}$, and $\|\cdot\|_p$ denote the transpose, conjugate transpose, matrix inversion, Moore-Penrose matrix inversion, and $l_p$ norm operation, respectively;
$|\Gamma|$ denotes the number of elements in set $\Gamma$ while $\|\mathbf{x}\|_0$ denotes the number of non-zero elements in vector $\mathbf{x}$;
$\mathbf{h}_{\Gamma}$ denotes the entries of the vector $\mathbf{h}$ in the set $\Gamma$;
$\mathbf{\Phi}_{\Gamma}$ denotes the submatrix comprising the $\Gamma$ columns of $\mathbf{\Phi}$.

\section{System Model}

In this paper, we consider the reconstruction of dynamic sparse channels $\mathbf{H}=[\mathbf{h}^{(1)}, \mathbf{h}^{(2)}, \cdots, \mathbf{h}^{(\tau)}]$, from under-sampled measurements $\mathbf{Y}=[\mathbf{y}^{(1)}, \mathbf{y}^{(2)}, \cdots, \mathbf{y}^{(\tau)}]$ in $\tau$ time slots.
The sparsity level, i.e., the maximum number of non-zero channel taps in $\mathbf{h}^{(t)}$ is $K$.
The measurement vector $\mathbf{y}^{(t)}=[y_1^{(t)}, y_2^{(t)}, \cdots, y_M^{(t)}]^T$ in the $t$-th time slot is usually obtained through a linear measurement process
\begin{equation}
\label{yt=phiht+nt}
\mathbf{y}^{(t)}=\mathbf{\Phi}\mathbf{h}^{(t)}+\mathbf{n}^{(t)},\hspace{0.3cm} t=1,2,\cdots,\tau,
\end{equation}
where $\mathbf{h}^{(t)}=[h_1^{(t)}, h_2^{(t)}, \cdots, h_N^{(t)}]^T$ denotes the dynamic sparse channel in the $t$-th time slot with $N>M$, $\mathbf{n}^{(t)}$ denotes the additive white Gaussian noise (AWGN) vector subject to the distribution $\mathcal{CN}(\mathbf{0},\mathbf{I}_M\sigma^2)$, and $\tau$ denotes the coherence time of the dynamic sparse channel.
The measurement matrix $\mathbf{\Phi}^{(t)}$ in (\ref{yt=phiht+nt}) is usually assumed to be time-invariant to simplify the design of communication systems \cite{ASOMP}:
\begin{equation}
\mathbf{\Phi}^{(t)}=\mathbf{\Phi}=[\boldsymbol{\phi}_1,\boldsymbol{\phi}_2, \cdots, \boldsymbol{\phi}_N], \hspace{0.1cm} t=1,2,\cdots,\tau.
\end{equation}
Since $\mathbf{\Phi}$ is usually designed in a random way, without loss of generality, we assume $\boldsymbol{\phi}_i^H\boldsymbol{\phi}_i=1$ and $\boldsymbol{\phi}_i^H\boldsymbol{\phi}_j\approx 0, \forall i\neq j$ \cite{CEhistory}.

\begin{figure}
\center{\includegraphics[width=0.45\textwidth]{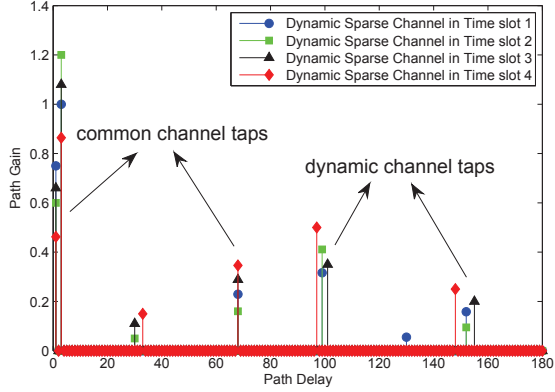}}
\caption{Illustration of the dynamic Vehicular B channel with a velocity of 100 km/h in four consecutive time slots.}
\label{Dynamic_Sparse_Channel}
\vspace{-0.4cm}
\end{figure}

The temporal correlations of practical wireless channels have been verified through analysis and experiments, even when the channels are varying fast \cite{ASOMP}.
Fig. \ref{Dynamic_Sparse_Channel} illustrates the time-domain impulse response of the dynamic Rayleigh fading Vehicular B channel with a velocity of 100km/h in four consecutive time slots, where each time slot corresponds to one OFDM symbol duration of 66.7 us in LTE-A systems \cite{VehicularB}.
It is clear that path delays of such dynamic sparse channels change slowly, while path gains evolve faster.
In order to characterize the temporal correlations of such dynamic sparse channels, we define a common path delay set as
\begin{equation}
\Gamma_0=\{i: i\in\Gamma^{(t)}, t=1,2,\cdots,\tau\},
\end{equation}
where $\Gamma^{(t)}=\{i: i\in\Omega, h_i^{(t)}\neq 0\}$ denotes the path delay set of $\mathbf{h}^{(t)}$ and $\Omega=\{1,2,\cdots,N\}$ denotes the entire set.
The size of the common path delay set is defined as $|\Gamma_0|=L\leq K$, where $L$ denotes the temporal correlation degree of dynamic sparse channels.
Thus we have
\begin{equation}
|\Gamma^{(i)}\setminus\Gamma^{(j)}|\leq K-L, \hspace{0.1cm} i\neq j,
\end{equation}
which means that the number of dynamic channel taps between any two time slots is constraint by $K-L$.

\section{The Proposed SMP Algorithm}
\vspace{0.1cm}
In this section, we propose the SMP algorithm to reconstruct dynamic sparse channels, whose path delays can be divided into two parts, i.e., the common channel taps and the dynamic channel taps.
The key idea of the proposed SMP algorithm is that, common channel taps in all $\tau$ time slots can be simultaneously obtained at first, and then the dynamic channel taps in each time slot can be individually detected with low complexity to refine the final reconstruction.
The pseudocode of the proposed SMP algorithm is provided in \textbf{Algorithm \ref{SMP}}, which is mainly comprised of the following two parts:

\vspace{0.1cm}
\begin{algorithm}[htb]
\renewcommand{\algorithmicrequire}{\textbf{Input:}}
\renewcommand\algorithmicensure {\textbf{Output:} }
\caption{Proposed SMP algorithm}
\label{SMP}
\begin{algorithmic}[1]
\REQUIRE ~~\\
Received signals: $\mathbf{Y}=[\mathbf{y}^{(1)},\mathbf{y}^{(2)},\cdots,\mathbf{y}^{(\tau)}]$;\\
Measurement matrix: $\mathbf{\Phi}=[\boldsymbol{\phi}_{1}, \boldsymbol{\phi}_{2}, \cdots, \boldsymbol{\phi}_{N}]$; \\
Maximum sparsity level: $K$.\\
\ENSURE ~~\\
Reconstructed channels: $\hat{\mathbf{H}}=[\hat{\mathbf{h}}^{(1)},\hat{\mathbf{h}}^{(2)},\cdots,\hat{\mathbf{h}}^{(\tau)}]$. \\
\STATE $\hat{\Gamma}_0=\varnothing$, $\mathbf{R}=\mathbf{Y}$.
\FOR {$k=1$ to $K$} 
\STATE $\mathbf{Z}=\mathbf{\Phi}^H\mathbf{R}$.
\STATE $\hat{\Gamma}_0=\hat{\Gamma}_0\cup \argmax_i \lambda_i=\sum_j|z_{i,j}|$.
\STATE $\mathbf{R}=\mathbf{Y}-\mathbf{\Phi}_{\hat{\Gamma}_0}\mathbf{\Phi}_{\hat{\Gamma}_0}^{\dagger}\mathbf{Y}$.
\ENDFOR
\FOR {$t=1$ to $\tau$}
\STATE $l = 1, \hat{\mathbf{h}}^{l}=\mathbf{\Phi}_{\hat{\Gamma}_0}^{\dagger}\mathbf{y}, \hat{\Gamma}^{l}=\hat{\Gamma}_0, \mathbf{r}^{l}=\mathbf{y}-\mathbf{\Phi}\hat{\mathbf{h}}^{l}.$
\WHILE {$l=1$ or $\|\mathbf{r}^{l}\|_2<\|\mathbf{r}^{l-1}\|_2$}
\STATE $l=l+1$.
\STATE $\tilde{\Gamma}^{l}=\hat{\Gamma}^{l-1}\cup \argmax_i \beta_i=\boldsymbol{\phi}_i^H\mathbf{r}^{l-1}$.
\STATE $\tilde{\mathbf{h}}^{l}=\mathbf{\Phi}_{\tilde{\Gamma}^{l}}^{\dagger}\mathbf{y}$.
\STATE $\hat{\Gamma}^{l}=\{K \text{ indices with largest magnitude in }\tilde{\mathbf{h}}^l\}$.
\STATE $\mathbf{r}^{l}=\mathbf{y}-\mathbf{\Phi}_{\hat{\Gamma}^{l}}\mathbf{\Phi}_{\hat{\Gamma}^{l}}^{\dagger}\mathbf{y}$.
\ENDWHILE
\STATE $\hat{\Gamma}=\hat{\Gamma}^{l-1}, \hat{\mathbf{h}}^{(t)}=\mathbf{\Phi}_{\hat{\Gamma}}^{\dagger}\mathbf{y}^{(t)}$.
\ENDFOR
\RETURN $\hat{\mathbf{H}}=[\hat{\mathbf{h}}^{(1)},\hat{\mathbf{h}}^{(2)},\cdots,\hat{\mathbf{h}}^{(\tau)}]$.
\end{algorithmic}
\end{algorithm}

\subsubsection{Common Channel Taps Detection (step 2$\sim$step 6)}
The first part of the proposed SMP algorithm aims to simultaneously estimate the common path delay set $\hat{\Gamma}_0$, which is expected to contain most of the true common channel taps.
Similar to our previously proposed A-SOMP algorithm \cite{ASOMP} and SSP algorithm \cite{SSP}, we detect the common channel taps by iteratively selecting the element that correlates best with the residue signal.
However, different from the simultaneous CS-based methods like A-SOMP \cite{ASOMP} and SSP \cite{SSP} which directly use $\hat{\Gamma}_0$ to reconstruct the dynamic sparse channel, the proposed SMP algorithm utilizes $\hat{\Gamma}_0$ as the initialization for the dynamic channel taps tracking.

\subsubsection{Dynamic Channel Taps Tracking (step 7$\sim$step 17)}
After the estimated common path delay set $\hat{\Gamma}_0$ has been obtained, the second part of the proposed SMP algorithm aims to track the dynamic channel taps, which are different from one time slot to another.
In each time slot (we omit the time slot superscript $t$ for notational simplicity), the estimated common path delay set is used as the initialization of the dynamic tracking process, whereby the path delay set $\hat{\Gamma}^l$ of size $K$ is maintained but refined in each iteration.
Specifically, a channel tap, which is considered reliable in one iteration but shown to be wrong in another, can be added into or removed from the path delay set $\hat{\Gamma}^l$.
Benefitting from the common channel taps detection, the dynamic channel taps tracking process is able to approach the true path delay set $\Gamma$ rapidly, and this process will be terminated when the residue error signal $\mathbf{r}^l$ is not reduced, i.e., $\|\mathbf{r}^l\|_2\geq \|\mathbf{r}^{l-1}\|_2$.
Finally, the dynamic sparse channel can be reconstructed based on the refined path delay set $\hat{\Gamma}$.

The proposed SMP algorithm is essentially different from the CS-based algorithms which ignore any temporal correlation of
dynamic sparse channels \cite{OMP}-\cite{SP}.
Meanwhile, it also differs from the simultaneous CS-based algorithms \cite{ASOMP}, \cite{SSP}, which assume the dynamic sparse channels in several consecutive time slots shares exactly the same path delay set.

\section{Performance Analysis}
To investigate the performance guarantee of the proposed SMP algorithm, in this section, we first derive the condition of the successful detection of the common channel taps, and then analyze the condition of the successful tracking of the dynamic channel taps.

A widely used condition of $\mathbf{\Phi}$ for reconstruction performance analysis is named RIP \cite{SP}, whereby a sensing matrix $\mathbf{\Phi}$ is said to satisfy the RIP of order $K$ if there exists a constant $\delta\in(0,1)$ such that
\begin{equation}
\label{RIPcondition}
(1-\delta)\|\mathbf{x}\|_2^2\leq\|\mathbf{\Phi x}\|_2^2\leq(1+\delta)\|\mathbf{x}\|_2^2
\end{equation}
holds true for all $K$-sparse vectors $\mathbf{x}\in\mathcal{R}^{N\times 1}$ ($\|\mathbf{x}\|_0=K$).
In particular, the minimum of all constant $\delta$ satisfying (\ref{RIPcondition}) is referred to as a restricted isometry constant $\delta_K$.

\subsection{Common Channel Taps Detection}
For the common channel taps detection, the proposed SMP algorithm selects the element that maximizes the sum of absolute correlation between the basis of $\mathbf{\Phi}$ and the residue measurement $\mathbf{R}$.
This greedy selection criterion can be formalized as follows:
\begin{equation}
\label{lambda}
\argmax_{\;\;\;\;\;\;\;\;i} \lambda_i=\sum_{t=1}^{\tau}|z_{i,t}|, \hspace{0.3cm} i=1,2,\cdots,N,
\end{equation}
where $z_{i,t}$ denotes the correlation between the $i$-th basis of $\mathbf{\Phi}$ and the residue measurement $\mathbf{r}^{(t)}$ in the $t$-th time slot.

In the first iteration, which means that the residue measurement $\mathbf{R}$ equals to $\mathbf{Y}$ and the rough common path delay set $\hat{\Gamma}_0$ is empty, the index $i$ is selected if and only if
\begin{equation}
\label{lambdajlambdai}
\lambda_i>\lambda_j, \hspace{0.3cm} \forall j\in\Omega \text{ and } j\neq i.
\end{equation}

Here we introduce two results of the RIP condition (\ref{RIPcondition}). The first one is named \textit{near-orthogonality} \cite{SP}: Let $I,J\subset\Omega$ be two disjoint sets, i.e., $I\cap J=\varnothing$, if $\delta_{|I|+|J|}<1$, then for an arbitrary vector $\mathbf{x}\in \mathcal{C}^{|J|}$, we have
\begin{equation}
\|\mathbf{\Phi}_I^H\mathbf{\Phi}_J\mathbf{x}\|_2\leq \delta_{|I|+|J|}\|\mathbf{x}\|_2.
\end{equation}
The second one is called \textit{RIP inequality} \cite{SP}: For an arbitrary vector $\mathbf{x}\in \mathcal{C}^{|I|}$ where $I\subset \Omega$, we have \\
\begin{equation}
(1-\delta_{|I|})\|\mathbf{x}\|_2\leq \|\mathbf{\Phi}_{I}^H\mathbf{\Phi}_{I}\mathbf{x}\|_2\leq(1+\delta_{|I|})\|\mathbf{x}\|_2,
\end{equation}
\begin{equation}
\frac{1}{1+\delta_{|I|}}\|\mathbf{x}\|_2\leq \|(\mathbf{\Phi}_I^H\mathbf{\Phi}_I)^{-1}\mathbf{x}\|_2\leq \frac{1}{1-\delta_{|I|}}\|\mathbf{x}\|_2.
\end{equation}

Then, by applying \textit{near-orthogonality} and \textit{RIP inequality}, $\lambda_i$ in (\ref{lambda}) can be derived as
\begin{eqnarray}
\label{lambdai}
\lambda_i&=&\sum_{t=1}^{\tau}|\boldsymbol{\phi}_i^H \mathbf{y}^{(t)}|=\sum_{t=1}^{\tau}|\boldsymbol{\phi}_i^H(\mathbf{\Phi}\mathbf{h}^{(t)}+\mathbf{n}^{(t)})|\nonumber\\
&\geq&\sum_{t=1}^{\tau}(|\boldsymbol{\phi}_i^H\boldsymbol{\phi}_i h_i^{(t)}|-\sum_{s\neq i}|\boldsymbol{\phi}_i^H\boldsymbol{\phi}_s h_s^{(t)}|-|\boldsymbol{\phi}_i^H\mathbf{n}^{(t)}|)\nonumber\\
&\geq&\sum_{t=1}^{\tau}(|h_i^{(t)}|-\sum_{s\neq i}\delta_2|h_s^{(t)}|-|\boldsymbol{\phi}_i^H\mathbf{n}^{(t)}|).
\end{eqnarray}
Similarly, $\lambda_j$ can be derived as
\begin{equation}
\label{lambdaj}
\lambda_j\leq \sum_{t=1}^{\tau}(|h_j^{(t)}|+\sum_{s\neq j}\delta_2|h_s^{(t)}|+|\boldsymbol{\phi}_j^H\mathbf{n}^{(t)}|).
\end{equation}
Using (\ref{lambdajlambdai}), (\ref{lambdai}), and (\ref{lambdaj}), we can obtain the sufficient condition under which the index $i$ will be selected:
\begin{equation}
\label{lambdajlambdai1}
\sum_{t=1}^{\tau}|h_i^{(t)}|\geq\sum_{t=1}^{\tau}|h_j^{(t)}|+\alpha_j,\hspace{0.15cm}\forall j\in\Omega\text{ and }j\neq i,
\end{equation}
where the parameter $\alpha_j$ is
\begin{equation}
\alpha_j = \sum_{t=1}^{\tau}(\sum_{s\neq j}\delta_2|h_s^{(t)}|+\sum_{s\neq i}\delta_2|h_s^{(t)}|+|\boldsymbol{\phi}_i^H\mathbf{n}^{(t)}|+|\boldsymbol{\phi}_j^H\mathbf{n}^{(t)}|).
\end{equation}

From (\ref{lambdajlambdai1}), it is clear that when the sum of the absolute values of a channel tap is larger than that of other channel taps plus a small value $\alpha_j$, the index of this channel tap will be added into the rough common path delay set $\hat{\Gamma}_0$.
The size of the rough common path delay set $\hat{\Gamma}_0$ is $K$, while there are only $L<K$ common channel taps.
Thus, the true $L$ common channel taps will be contained in $\hat{\Gamma}_0$ with a great probability.
As shown in Fig. \ref{Dynamic_Sparse_Channel}, the sum of the absolute values of a common channel tap is much larger than that of the dynamic channel taps.
This is because the absolute values of a common channel tap in different time slots can be summed together while not true for the dynamic channel taps.

\subsection{Dynamic Channel Taps Tracking}
Intuitively, in each time slot (we omit the time slot superscript $t$ for notational simplicity), the proposed SMP algorithm tracks the dynamic channel taps to refine the final reconstruction, whereby a reliable channel tap will be added into and an unreliable one will be removed from the estimated path delay set $\hat{\mathbf{\Gamma}}$ in each iteration.
In order to measure the performance of the dynamic channel taps tracking process, the reconstruction distortion $\|\hat{\mathbf{h}}-\mathbf{h}\|_2$ is important to be analyzed.

Here we introduce the \textit{residue-orthogonality} \cite{SP} of the projection operator used in the CS theory: For an arbitrary vector $\mathbf{x}\in\mathcal{C}^{M}$ and a sampling matrix $\mathbf{\Phi}_I\in\mathcal{C}^{M\times |I|}$ of full column rank, we have
\begin{equation}
\left\{
  \begin{array}{ll}
    \mathbf{\Phi}_I^H\text{resid}(\mathbf{x},\mathbf{\Phi}_I)=\mathbf{0}, \\
    \|\text{resid}(\mathbf{x},\mathbf{\Phi}_I)\|_2\leq\|\mathbf{x}\|_2,
  \end{array}
\right.
\end{equation}
where $\text{resid}(\mathbf{x}, \mathbf{\Phi}_I)=\mathbf{x}-\mathbf{\Phi}_I\mathbf{\Phi}_I^{\dagger}\mathbf{x}$ denotes the residue signal.

Thus, by applying RIP condition and \textit{residue-orthogonality}, $\|\mathbf{r}^{l}\|_2$ of the proposed SMP algorithm can be derived as
\begin{eqnarray}
\label{rl}
\|\mathbf{r}^{l}\|_2&=&\|\text{resid}(\mathbf{y},\mathbf{\Phi}_{\hat{\Gamma}^l})\|_2 \nonumber\\
&\leq&\|\text{resid}(\mathbf{\Phi}_{\Gamma\setminus\hat{\Gamma}^l}\mathbf{h}_{\Gamma\setminus\hat{\Gamma}^l}, \mathbf{\Phi}_{\hat{\Gamma}^l})\|_2+\|\text{resid}(\mathbf{n},\mathbf{\Phi}_{\hat{\Gamma}^l})\|_2 \nonumber\\
&\leq&\|\mathbf{\Phi}_{\Gamma\setminus\hat{\Gamma}^l}\mathbf{h}_{\Gamma\setminus\hat{\Gamma}^l}\|_2+\|\mathbf{n}\|_2 \nonumber\\
&\leq& \sqrt{1+\delta_{2K+1}}\|\mathbf{h}_{\Gamma\setminus\hat{\Gamma}^l}\|_2+\|\mathbf{n}\|_2.
\end{eqnarray}
Similarly, $\|\mathbf{r}^{l-1}\|_2$ can be also derived as
\begin{eqnarray}
\label{rl1}
\|\mathbf{r}^{l-1}\|_2&=&\|\text{resid}(\mathbf{y},\mathbf{\Phi}_{\hat{\Gamma}^{l-1}})\|_2 \nonumber\\
&\geq&\frac{1-2\delta_{2K+1}}{\sqrt{1-\delta_{2K+1}}}\|\mathbf{h}_{\Gamma\setminus\hat{\Gamma}^{l-1}}\|_2-\|\mathbf{n}\|_2.
\end{eqnarray}

In order to compare $\|\mathbf{r}^l\|_2$ and $\|\mathbf{r}^{l-1}\|_2$, we need to obtain the relationship between $\|\mathbf{h}_{\Gamma\setminus\hat{\Gamma}^l}\|_2$ and $\|\mathbf{h}_{\Gamma\setminus\hat{\Gamma}^{l-1}}\|_2$.
From the set $\tilde{\Gamma}^l$ in the process of dynamic channel taps tracking, we can derive the following two inequalities
\begin{equation}
\label{18}
\frac{\|\mathbf{h}_{\Gamma\setminus\tilde{\Gamma}^l}\|_2}{\sqrt{K}}\leq\frac{2\delta_{2K+1}}{(1-\delta_{2K+1})^2}\|\mathbf{h}_{\Gamma\setminus\hat{\Gamma}^{l-1}}\|_2+\frac{2(1+\delta_{2K+1})}{1-\delta_{2K+1}}\|\mathbf{n}\|_2,
\end{equation}
\begin{equation}
\label{19}
\|\mathbf{h}_{\Gamma\setminus\hat{\Gamma}^l}\|_2\leq\frac{1+\delta_{2K+1}}{1-\delta_{2K+1}}\|\mathbf{h}_{\Gamma\setminus\tilde{\Gamma}^l}\|_2+\frac{2}{1-\delta_{2K+1}}\|\mathbf{n}\|_2.
\end{equation}
The proof of these two inequalities is similar to that of Appendices H and I in \cite{SP}, and we omit the proof here due to their similarity and paper space limitation.
By applying (\ref{18}) and (\ref{19}) into (\ref{rl}) and (\ref{rl1}), the termination condition $\|\mathbf{r}^{l}\|_2\geq\|\mathbf{r}^{l-1}\|_2$ for dynamic channel taps tracking can lead to the result as
\begin{equation}
\label{bsss}
\|\mathbf{n}\|_2\geq\frac{\delta_{2K+1}}{\sqrt{K}+2}\|\mathbf{h}_{\Gamma\setminus\hat{\Gamma}^{l-1}}\|_2=\frac{\delta_{2K+1}}{\sqrt{K}+2}\|\mathbf{h}_{\Gamma\setminus\hat{\Gamma}}\|_2,
\end{equation}
which means that if the noise power is larger than that of the residue channel taps multiplied with a coefficient, the dynamic tracking process will be terminated, and the final estimated path delay set can be obtained as $\hat{\Gamma}=\hat{\Gamma}^{l-1}$.

Then we analyze the relationship between the reconstruction distortion $\|\hat{\mathbf{h}}-\mathbf{h}\|_2$ and $\|\mathbf{h}_{\Gamma\setminus\hat{\Gamma}}\|_2$ as
\begin{eqnarray}
\label{asss}
\|\hat{\mathbf{h}}-\mathbf{h}\|_2&\leq&\|\mathbf{h}_{\hat{\Gamma}}-\mathbf{\Phi}_{\hat{\Gamma}}^{\dagger}\mathbf{y}\|_2+\|\mathbf{h}_{\Gamma\setminus\hat{\Gamma}}\|_2\nonumber\\
&\leq&\|\mathbf{h}_{\hat{\Gamma}}-\mathbf{\Phi}_{\hat{\Gamma}}^{\dagger}(\mathbf{\Phi}_{\Gamma}\mathbf{h}_{\Gamma})\|_2+\|\mathbf{\Phi}_{\hat{\Gamma}}^{\dagger}\mathbf{n}\|_2+\|\mathbf{h}_{\Gamma\setminus\hat{\Gamma}}\|_2\nonumber\\
&\leq&\frac{1}{1-\delta_{2K+1}}\|\mathbf{h}_{\Gamma\setminus\hat{\Gamma}}\|_2+\frac{1+\delta_{2K+1}}{1-\delta_{2K+1}}\|\mathbf{n}\|_2,
\end{eqnarray}
where $\hat{\mathbf{h}}_{\hat{\Gamma}}=\mathbf{\Phi}_{\hat{\Gamma}}\mathbf{\Phi}_{\hat{\Gamma}}^{\dagger}\mathbf{y}$ and $\hat{\mathbf{h}}_{\Omega\setminus\hat{\Gamma}}=\mathbf{0}$.
By substituting (\ref{asss}) into (\ref{bsss}), we have
\begin{equation}
\|\hat{\mathbf{h}}-\mathbf{h}\|_2\leq\frac{2+\sqrt{K}+\delta_{2K+1}+\delta_{2K+1}^2}{\delta_{2K+1}(1-\delta_{2K+1})}\|\mathbf{n}\|_2,
\end{equation}
which means that the reconstruction distortion is linearly upper bounded by the noise power $\|\mathbf{n}\|_2$, the sparsity level $K$, and the RIP of the measurement matrix $\mathbf{\Phi}$.

In addition, the main computational burden of the proposed SMP algorithm comes from the matrix pseudo inversion calculating, which can be effectively solved by the Gram-Schmidt algorithm \cite{GS}.
Compared with the conventional reconstruction algorithms, the complexity of the proposed SMP algorithm is competitive.

\section{Simulation Results}
In this section, we investigate the performance of the proposed SMP algorithm through numerical simulations.
We construct the measurement matrix $\mathbf{\Phi}$ of size $M\times N=100\times 200$ whose elements are independent and identically distributed (i.i.d.), i.e., $\phi_{i,j}\sim\mathcal{N}(0,1/M)$.
The gains of channel taps at each time slot are drawn from the standard Gaussian distribution, i.e., $h_i^{(t)}\sim\mathcal{N}(0,1), i\in\Gamma^{(t)}$, while the path delays of the dynamic channel taps for the $t$-th time slot are randomly chosen from $\Omega\setminus\Gamma_0$.
The coherence time of the dynamic sparse channel is set as $\tau=10$.

A commonly used metric to evaluate the channel reconstruction performance in practical communication systems is the mean squared error (MSE) defined as
\begin{equation}
\text{MSE} = \frac{1}{\tau}\sum_{t=1}^{\tau}\frac{\|\hat{\mathbf{h}}^{(t)}-\mathbf{h}^{(t)}\|_2}{\|\mathbf{h}^{(t)}\|_2}.
\end{equation}
In the simulation, we perform $10,000$ independent trials to obtain the MSE performance of the following reconstruction algorithms: 1) Linear method \cite{CEhistory}; 2) OMP \cite{OMP}; 3) SP \cite{SP}; 4) A-SOMP \cite{ASOMP}; 5) Oracle least square (oracle-LS) \cite{OracleLS}.
Specifically, the oracle-LS is regarded as the theoretical bound of the MSE performance due to the fact that the path delay set is assumed to be perfectly known as a prior information.

\begin{figure}
\center{\includegraphics[width=0.44\textwidth]{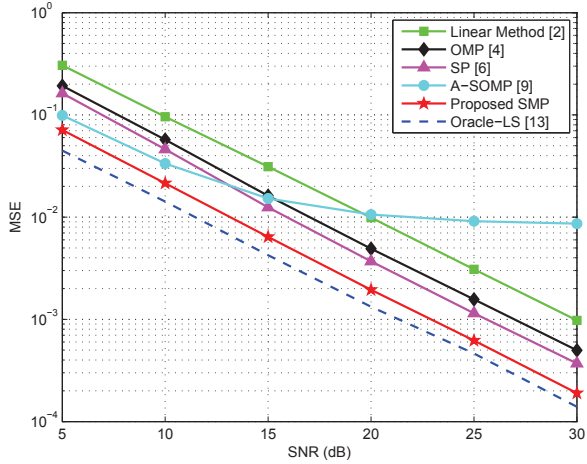}}
\caption{MSE performance comparison against the SNR.}
\label{MSE_SNR}
\end{figure}

Fig. \ref{MSE_SNR} investigates the MSE performance comparison against SNR for different reconstruction algorithms, where the temporal correlation degree is set as $L=5$.
It is evident that the standard OMP algorithm and the SP algorithm outperform the linear method by about 2 dB, where the benefit comes from utilizing the channel sparsity.
Further, the A-SOMP algorithm is better than the SP algorithm by about 2 dB when the SNR is low, since it roughly considers the temporal correlations of the dynamic sparse channel.
For the proposed SMP algorithm, it is clear that another 2 dB SNR gain can be achieved due to its capability to detect the common channel taps accurately and track the dynamic channel taps rapidly as discussed in Section IV.

\begin{figure}
\center{\includegraphics[width=0.44\textwidth]{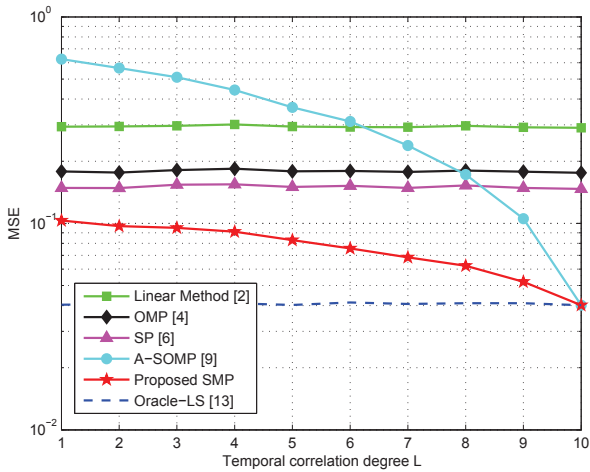}}
\vspace{-0.2cm}
\caption{MSE performance comparison against the temporal correlation degree.}
\label{MSE_L}
\vspace{-0.4cm}
\end{figure}

Fig. \ref{MSE_L} shows the MSE performance comparison against the temporal correlation degree $L$, where $\text{SNR}=5$ dB is considered.
It is clear that the SP algorithm achieves lower error levels than the standard OMP algorithm, while the conventional linear method performs worst.
The MSE performance of the A-SOMP algorithm is competitive only when the temporal correlation degree $L$ is very large, while the MSE performance of the proposed SMP algorithm is the best for all considered temporal correlation degree.
It should be pointed out that when there are no dynamic channel taps, i.e., $L=K$, the MSE performance of the proposed SMP algorithm and the A-SOMP algorithm achieves the performance bound of the oracle-LS method.

\balance
\section{Conclusions}
In this paper, we proposed a structured CS algorithm called SMP to reconstruct dynamic sparse channels for broadband wireless communication systems.
By exploiting the temporal correlations of dynamic sparse channels, the path delays are divided into two parts, i.e., the common channel taps and the dynamic channel taps.
Based on this separation, the common channel taps over all time slots are simultaneously detected at first.
Then, the proposed SMP algorithm tracks the dynamic channel taps for each single time slot individually.
The theoretical analysis of the proposed SMP algorithm provides the condition of successfully detecting the common channel taps as well as the upper bound of the reconstruction distortion.
Finally, simulation results demonstrated that the proposed SMP algorithm is able to accurately reconstruct dynamic sparse channels with about 2 dB SNR gain compared with the recently proposed simultaneous CS-based reconstruction algorithms.

\section*{Acknowledgment}
This work was supported by National Key Basic Research Program of China (Grant No. 2013CB329203), grant-in-aid for the Japan Society for the Promotion of Science (JSPS) fellows (Grant No. 2402366), and National Natural Science Foundation of China (Grant Nos. 61271266 and 61411130156).

\end{document}